\documentclass[aps,amssymb,prd,reprint,groupedaddress]{revtex4-1}

\usepackage{amsmath}
\usepackage{graphicx}
\usepackage{epsfig}

\newcommand{\nn}{\nonumber}

\begin{document}


\title{Bounding the Greybody Factors for Non-rotating Black Holes}


\author{Tritos Ngampitipan}
\email{tritos.ngampitipan@gmail.com}
\affiliation{Department of Physics, Faculty of Science,\\
             Chulalongkorn University, Bangkok 10330, Thailand}

\author{Petarpa Boonserm}
\email{petarpa.boonserm@gmail.com}
\affiliation{Department of Mathematics and Computer Science, Faculty of Science,\\
             Chulalongkorn University, Bangkok 10330, Thailand}


\date{\today}

\begin{abstract}
Semiclassical black holes emit radiation called Hawking radiation. Such radiation, as seen by an asymptotic observer far outside the black hole, differs from original radiation near the horizon of the black hole by a redshift factor and the so-called ``greybody factor". In this paper, we concentrate on the greybody factor-various bounds for the greybody factors of non-rotating black holes are obtained, concentrating on charged Reissner-Nordstr\"{o}m and Reissner-Nordstr\"{o}m-de Sitter black holes. These bounds can be derived by using a $2 \times 2$ transfer matrix formalism. It is found that the charges of black holes act as efficient barriers. Furthermore, adding extra dimensions to spacetime can shield Hawking radiation. Finally, the cosmological constant can increase the emission rate of Hawking radiation.
\end{abstract}


\maketitle

\section{Introduction}
Classically a  black hole is associated with the concept that anything which enters the black hole cannot escape. In 1974, Stephen Hawking, however, showed that semi-classically a black hole could indeed emit quantum radiation, an effect which became known as Hawking radiation \cite{Hawking}. This effect was derived by studying quantum field theory in a black hole background. In the context of quantum field theory, creation and annihilation of particles are possible. If pair production occurs near a black hole horizon, one can picture Hawking radiation as one of the particles from pair production falling in, with the other moving away from the black hole. An observer outside the black hole can see this particle as Hawking radiation. But according to general relativity, a black hole curves spacetime around it. This nontrivial spacetime behaves as gravitational potential under which particles move. Some of them are reflected back into the black hole and others are transmitted out of the black hole. Therefore, Hawking radiation seen by an observer far outside the black hole differs from radiation which had not yet been scattered by the gravitational potential. This difference can be measured by the so-called ``greybody factor".

There have been a number of studies devoted to calculating these greybody factors. Some used the WKB approximation to calculate the greybody factors of the four-dimensional Schwarzschild and Reissner-Nordstr\"{o}m black holes \cite{Parikh, Fleming, Lange}. Some solved the wave equation in a black hole background by various approximations \cite{Fernando, Kim, esc}. However, there is a rather different analytic technique to derive rigorous bounds on the greybody factors \cite{Visser, Bogoliubov, PhD thesis}. By using this method, bounds on the greybody factors of the four-dimensional Schwarzschild black holes was obtained in Ref. \cite{pet}. In this paper, we extend the analysis and derive rigorous bounds for the greybody factors of the four- dimensional Reissner-Nordstr\"{o}m black holes, the higher dimensional Schwarzschild-Tangherlini black holes, the charged dilatonic black holes in (2 + 1) dimensions, and the charged dilatonic black holes in (3 + 1) dimensions.

\section{The Reissner-Nordstr\"{o}m black holes}
The Reissner-Nordstr\"{o}m (RN) metric is given by
\begin{equation}
ds^{2} = -\Delta dt^{2} + \Delta^{-1}dr^{2} + r^{2}d\Omega^{2},
\end{equation}
where $d\Omega^{2} = d\theta^{2} + \sin^{2}\theta d\phi^{2}$ and
\begin{equation}
\Delta = 1 - \frac{2GM}{r} + \frac{G\left(Q^{2} + P^{2}\right)}{r^{2}}.
\end{equation}
The Schr\"{o}dinger-like equation governing the modes is given by
\begin{equation}
\frac{d^{2}\psi}{dr_{*}^{2}} + \left[\omega^{2} - V(r)\right]\psi = 0,
\end{equation}
where $r_{*}$ is the standard ``tortoise coordinate"
\begin{equation}
dr_{*} = \frac{1}{\Delta}dr.
\end{equation}
and
\begin{equation}
V(r) = \frac{l(l + 1)\Delta}{r^{2}} + \frac{\Delta\partial_{r}\Delta}{r}.\label{V(r)}
\end{equation}

\begin{figure}[pb]
\centerline{\psfig{file=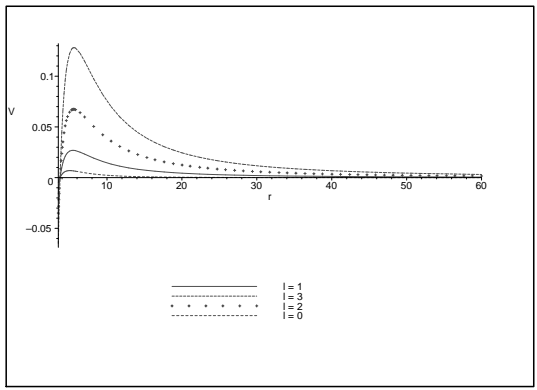,width=2in, angle = -90}}
\vspace*{1pt}
\caption{The Reissner-Nordstr\"{o}m potential with $Q = 1$ and $M = 2$ in different angular momenta.}\label{RN pot}
\end{figure}

\noindent Some purely technical computations are relegated to Appendix A. We can see the structure of the Reissner-Nordstr\"{o}m potential with $Q = 1$ and $M = 2$ from Fig. \ref{RN pot}.

Using the analysis of \cite{Visser, Bogoliubov, PhD thesis}, lower bounds on the transmission probabilities are given by
\begin{equation}
T \geq \text{sech}^{2}\left(\int_{-\infty}^{\infty}\vartheta dr_{*}\right),
\end{equation}
where
\begin{equation}
\vartheta = \frac{\sqrt{(h')^{2} + \left(\omega^{2} - V - h^{2}\right)^{2}}}{2h},
\end{equation}
for some positive function $h$. We set $h = \omega$, then
\begin{eqnarray}
T &\geq& \text{sech}^{2}\left(\frac{1}{2\omega}\int_{-\infty}^{\infty}Vdr_{*}\right)\nn\\
  &=&    \text{sech}^{2}\left[\frac{1}{2\omega}\left\{\frac{l(l + 1)}{GM + A} + \frac{GM + 2A}{3(GM + A)^{2}}\right\}\right],\label{TRN bound}
\end{eqnarray}
where
\begin{equation}
A^{2} \equiv G^{2}M^{2} - G\left(Q^{2} + P^{2}\right).
\end{equation}

\begin{figure}[pb]
\centerline{\psfig{file=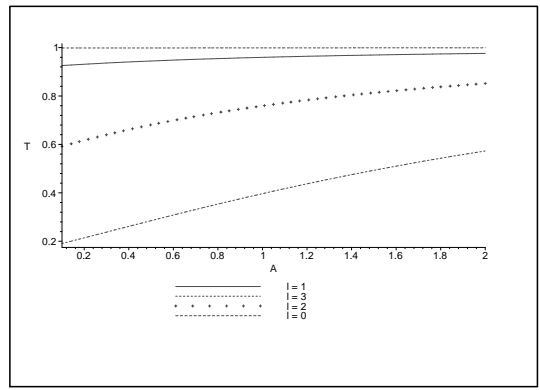,width=2in, angle = -90}}
\vspace*{1pt}
\caption{Dependence of the bound of the greybody factor on the Reissner-Nordstr\"{o}m black hole charges in different angular momenta.}\label{TRN2}
\end{figure}

\noindent If the black holes have no electric charges or magnetic charges, it is found that $A = GM$ and the above bound is reduced to
\begin{equation}
T \geq \text{sech}^{2}\left[\frac{2l(l + 1) + 1}{8GM\omega}\right],
\end{equation}
which is exactly the bound for the Schwarzschild black holes emitting spinless particles \cite{pet}. From Fig. \ref{TRN2}, the graph is plotted by setting $GM = 2$ and $\omega = 2$. The point $A = 2$ corresponds to the uncharged RN black hole (which is the Schwarzschild black hole). The point $A < 2$ describes the effects of charges on the bound of the greybody factor. From the value of $A$, decreasing of $A$ corresponds to increasing the magnitude of charges. The graph shows that when the magnitude of charges increases, the bound of the greybody factor decreases. That is the charges are good barriers to resist tunneling of uncharged scalar particles. Moreover, the transmission coefficients is smaller in higher angular momenta.

By using the WKB approximation, the approximate transmission coefficient is given by \cite{Lange}
\begin{equation}
T \approx T_{\text{WKB}} = \exp\left[-\frac{2}{\hbar}\text{Im}\int_{a}^{b}p(x)dx\right],
\end{equation}
where
\begin{equation}
p(x) = \sqrt{2m[E - V(x)]}.
\end{equation}
We find that
\begin{eqnarray}
T_{\text{WKB}} &=& \exp\left[-\frac{2\pi}{\hbar}\left\{2G\omega\left(M - \frac{\omega}{2}\right)\right.\right.\nn\\
  && - (M - \omega)\sqrt{G^{2}\left(M - \omega\right)^{2} - G\left(Q^{2} + P^{2}\right)}\nn\\
  && \left.\left. + M\sqrt{G^{2}M^{2} - G\left(Q^{2} + P^{2}\right)}\right\}\right].\label{Twkb RN}
\end{eqnarray}
Derivation of this equation is given in Appendix B. Turning now to an asymptotic analysis inspired by studies of quasi-normal modes, the approximate transmission coefficient for large $\omega$ is given by \cite{Neitzke, Nozari, Zhai}
\begin{equation}
T \approx T_{\text{asymptotic}} = \frac{e^{\beta\omega} - 1}{e^{\beta\omega} + 2 + 3e^{-\beta_{I}\omega}},\label{T asymp}
\end{equation}
where
\begin{eqnarray}
\beta     &=& \frac{8\pi M}{1 + Q^{2}/2GM^{2} + 5Q^{4}/16G^{2}M^{4}},\nn\\
\beta_{I} &=& -\frac{2\pi\left[GM - \sqrt{G^{2}M^{2} - GQ^{2}}\right]^{2}}{\sqrt{G^{2}M^{2} - GQ^{2}}}.
\end{eqnarray}

\begin{figure}[pb]
\centerline{\psfig{file=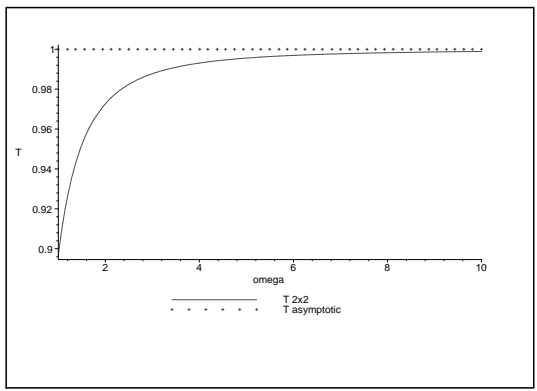,width=2in, angle = -90}}
\vspace*{1pt}
\caption{Comparison of the bound of the greybody factor of the Reissner-Nordstr\"{o}m black hole from the $2 \times 2$ transfer matrix and the asymptotic result.}\label{TRNwkb}
\end{figure}

\noindent The greybody factors obtained from the $2 \times 2$ transfer matrix formalism [Eq. (\ref{TRN bound})] are compared with the asymptotic result [Eq. (\ref{T asymp})] on the graph shown in Fig. \ref{TRNwkb}. The graph shows that the result from the $2 \times 2$ transfer matrix is closed to the asymptotic result at large $\omega$. Moreover, the $2 \times 2$ transfer matrix gives a true lower bound.

\section{The Schwarzschild-Tangherlini black holes}
The Schwarzschild-Tangherlini metric in $d$ dimensions is given by \cite{esc}
\begin{equation}
ds^{2} = -\left[1 - \left(\frac{r_{0}}{r}\right)^{d - 3}\right]dt^{2} + \left[1 - \left(\frac{r_{0}}{r}\right)^{d - 3}\right]^{-1}dr^{2} + r^{2}d\Omega^{2}_{d - 2},
\end{equation}
where the Schwarzschild radius $r_{0}$ in $d$ dimensions is given by
\begin{equation}
r_{0} = \frac{16\pi GM}{(d - 2)\Omega_{d - 2}},
\end{equation}
with
\begin{equation}
\Omega_{d - 2} = \frac{2\pi^{(d - 1)/2}}{\Gamma\left(\dfrac{d - 1}{2}\right)}.
\end{equation}

\begin{figure}[pb]
\centerline{\psfig{file=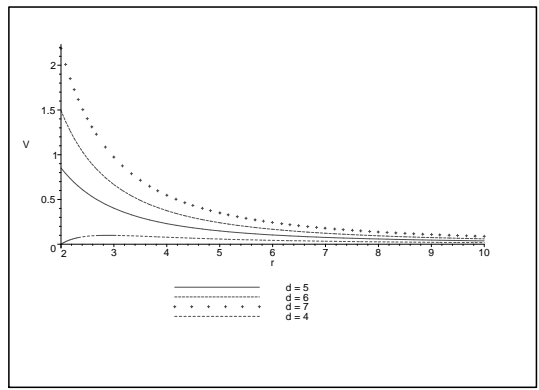,width=2.0in, angle = -90}}
\vspace*{1pt}
\caption{The higher dimensional potential with $l = 1$ and $GM = 1$ in various dimensions.}\label{hd pot}
\end{figure}

\noindent The Schr\"{o}dinger like equation is given by
\begin{equation}
\left[\frac{d^{2}}{dr_{*}^{2}} + \omega^{2} - V(r)\right]r^{(d - 2)/2}\varphi = 0,
\end{equation}
where
\begin{equation}
dr_{*} = \frac{1}{f(r)}dr
\end{equation}
and
\begin{eqnarray}
V(r) &=& \frac{(d - 2)(d - 4)}{4}\frac{f^{2}(r)}{r^{2}} + \frac{(d - 2)}{2}\frac{f(r)\partial_{r}f(r)}{r}\nn\\
     && + l(l + d - 3)\frac{f(r)}{r^{2}},
\end{eqnarray}
with
\begin{equation}
f(r) = 1 - \left(\frac{r_{0}}{r}\right)^{d - 3}.
\end{equation}
From Fig. \ref{hd pot}, the Schwarzschild-Tangherlini potential is plotted with $l = 1$ and $GM = 1$ in various dimensions.

The lower bound on the transmission probability for $h = \omega$ is
\begin{eqnarray}
T &\geq& \text{sech}^{2}\left(\frac{1}{2\omega}\int_{-\infty}^{\infty}Vdr_{*}\right)\nn\\
  &=&    \text{sech}^{2}\left[\frac{1}{2\omega}\int_{r_{0}}^{\infty}\left\{\frac{(d - 2)(d - 4)}{4}\frac{f(r)}{r^{2}}\right.\right.\nn\\
  && \left.\left. + \frac{(d - 2)}{2}\frac{\partial_{r}f(r)}{r} + \frac{l(l + d - 3)}{r^{2}}\right\}dr\right]\nn\\
  &=&    \text{sech}^{2}\left[\frac{(d - 2)(d - 3) + 4l(l + d - 3)}{8\omega r_{0}}\right].
\end{eqnarray}

\begin{figure}[pb]
\centerline{\psfig{file=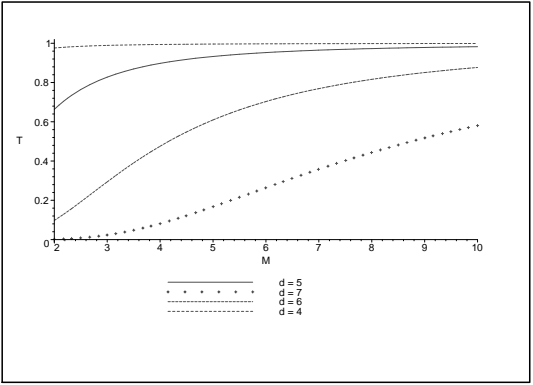,width=2in, angle = -90}}
\vspace*{1pt}
\caption{Dependence of the bound of the greybody factor on the black hole mass in various dimensions.}\label{ThdSM}
\end{figure}

\noindent If $d = 4$, this bound is reduced to
\begin{equation}
T \geq \text{sech}^{2}\left[\frac{2l(l + 1) + 1}{8GM\omega}\right],
\end{equation}
which is, again, exactly the bound for the four-dimensional Schwarzschild black holes emitting spinless particles. Figure \ref{ThdSM} shows the plot between the transmission coefficients and the black hole mass in various dimensions. The graph is plotted by setting $l = 1$ and $\omega = 2$. The line $d = 4$ corresponds to the four-dimensional Schwarzschild black hole. The graph shows that when the black hole mass increases, the bound of the greybody factor also increases. However, for the same mass the bound of the greybody factor is less in higher dimensions \cite{Nicolini}.

\section{The Charged dilatonic black holes in (2 + 1) dimensions}
The Charged Dilatonic metric in (2 + 1) dimensions is given by \cite{Fernando}
\begin{equation}
ds^{2} = -f(r)dt^{2} + \frac{4r^{2}}{f(r)}dr^{2} + r^{2}d\theta^{2},
\end{equation}
where
\begin{equation}
f(r) = -2Mr + 8\Lambda r^{2} + 8Q^{2}.
\end{equation}
For $M > 8Q\sqrt{\Lambda}$, this spacetime describes a black hole with two event horizons
\begin{equation}
r_{\pm} = \frac{M \pm \sqrt{M^{2} - 64Q^{2}\Lambda}}{8\Lambda}.
\end{equation}
The Schr\"{o}dinger like equation is given by
\begin{equation}
\left[\frac{d^{2}}{dr_{*}^{2}} + \omega^{2} - V(r)\right]u(r) = 0,
\end{equation}
where
\begin{equation}
dr_{*} = \frac{2r}{f(r)}dr
\end{equation}
and
\begin{eqnarray}
V(r) &=& -(8m^{2}\Lambda + 6m\Lambda) + 14\Lambda^{2}r\nn\\
     && + \left(\frac{5M^{2}}{8} + 2m^{2}M\right)\frac{1}{r}\nn\\
     && - (4MQ^{2} + 8m^{2}Q^{2})\frac{1}{r^{2}} + \frac{6Q^{4}}{r^{3}}.
\end{eqnarray}

\begin{figure}[pb]
\centerline{\psfig{file=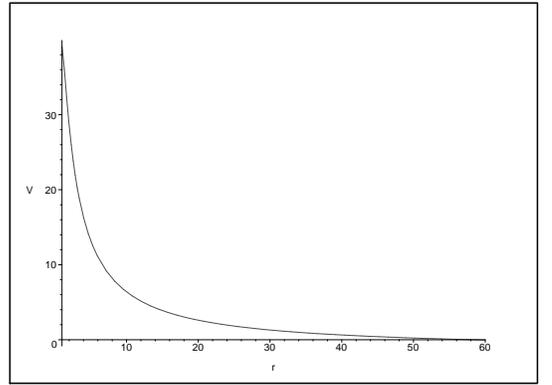,width=2.0in, angle = -90}}
\vspace*{1pt}
\caption{The (2 + 1) charged dilatonic potential with $m = 1$, $\Lambda = 0.1$, $Q = 1$, and $M = 10$.}\label{di2 pot}
\end{figure}

\noindent For simplicity, we consider the case of linearized cosmological constant. By linearized cosmological constant we mean it is so small that its higher power can be approximated by zero. The potential becomes
\begin{eqnarray}
V(r) &=& -(8m^{2}\Lambda + 6m\Lambda) + \left(\frac{5M^{2}}{8} + 2m^{2}M\right)\frac{1}{r}\nn\\
     && - (4MQ^{2} + 8m^{2}Q^{2})\frac{1}{r^{2}} + \frac{6Q^{4}}{r^{3}}.
\end{eqnarray}
The (2 + 1) charged dilatonic potential is plotted with $m = 1$, $\Lambda = 0.1$, $Q = 1$, and $M = 10$ as shown in Fig. \ref{di2 pot}. The coordinate $r_{*}$ can explicitly be written as
\begin{equation}
r_{*} = \frac{1}{4\Lambda(r_{+} - r_{-})}\left[r_{+}\ln(r - r_{+}) - r_{-}\ln(r - r_{-})\right].
\end{equation}
The lower bound on the transmission probability for $h = \omega$ is
\begin{eqnarray}
T &\geq& \text{sech}^{2}\left[\frac{1}{2\omega}\int_{-\infty}^{\infty}Vdr_{*}\right]\nn\\
  &=&    \text{sech}^{2}\left[\frac{1}{2\omega}\int_{r_{+}}^{\infty}\left\{-(8m^{2}\Lambda + 6m\Lambda)\right.\right.\nn\\
  && + \left(\frac{5M^{2}}{8} + 2m^{2}M\right)\frac{1}{r}\nn\\
  &&\left.\left. - (4MQ^{2} + 8m^{2}Q^{2})\frac{1}{r^{2}} + \frac{6Q^{4}}{r^{3}}\right\}\frac{2r}{f(r)}dr\right]\nn\\
  &=&    \text{sech}^{2}\left[-\frac{1}{\omega}\frac{272m\Lambda(4m + 3)}{15\sqrt{M^{2} - 64Q^{2}\Lambda}} + \frac{1}{\omega}\frac{11M\left(5M + 16m^{2}\right)}{96\sqrt{M^{2} - 64Q^{2}\Lambda}}\right.\nn\\
  && - \frac{1}{\omega}(M + 2m^{2})\left\{\frac{23}{15} - \frac{r_{+} - 1}{r_{+} + 1} - \frac{1}{3}\left(\frac{r_{+} - 1}{r_{+} + 1}\right)^{3}\right.\nn\\
  && \left. - \frac{1}{5}\left(\frac{r_{+} - 1}{r_{+} + 1}\right)^{5}\right\} + \frac{1}{\omega}\frac{3M}{16}\left\{\frac{23}{15} - \frac{r_{+} - 1}{r_{+} + 1}\right.\nn\\
  && \left. - \frac{1}{3}\left(\frac{r_{+} - 1}{r_{+} + 1}\right)^{3} - \frac{1}{5}\left(\frac{r_{+} - 1}{r_{+} + 1}\right)^{5}\right\}\nn\\
  && \left. + \frac{1}{\omega}\frac{6\Lambda Q^{2}}{M + \sqrt{M^{2} - 64Q^{2}\Lambda}}\right],\label{T 2+1}
\end{eqnarray}
where
\begin{eqnarray}
r_{+} - 1                   &=& \frac{M + \sqrt{M^{2} - 64Q^{2}\Lambda}}{8\Lambda} - 1\nn\\
                            &=& \frac{M + \sqrt{M^{2} - 64Q^{2}\Lambda} - 8\Lambda}{8\Lambda}\nn\\
r_{+} + 1                   &=& \frac{M + \sqrt{M^{2} - 64Q^{2}\Lambda}}{8\Lambda} + 1\nn\\
                            &=& \frac{M + \sqrt{M^{2} - 64Q^{2}\Lambda} + 8\Lambda}{8\Lambda}\nn\\
\frac{r_{+} - 1}{r_{+} + 1} &=& \frac{M + \sqrt{M^{2} - 64Q^{2}\Lambda} - 8\Lambda}{M + \sqrt{M^{2} - 64Q^{2}\Lambda} + 8\Lambda}.
\end{eqnarray}
The exact transmission coefficient is given by \cite{Fernando}
\begin{equation}
T = 1 - \frac{\cosh^{2}\left[\dfrac{\pi\omega}{4\Lambda} - \dfrac{\pi}{2}\sqrt{\dfrac{\omega^{2} - 8m^{2}\Lambda}{4\Lambda^{2}} - 1}\right]}{\cosh^{2}\left[\dfrac{\pi\omega}{4\Lambda} + \dfrac{\pi}{2}\sqrt{\dfrac{\omega^{2} - 8m^{2}\Lambda}{4\Lambda^{2}} - 1}\right]}.\label{T 2+1 paper}
\end{equation}

\begin{figure}[pb]
\centerline{\psfig{file=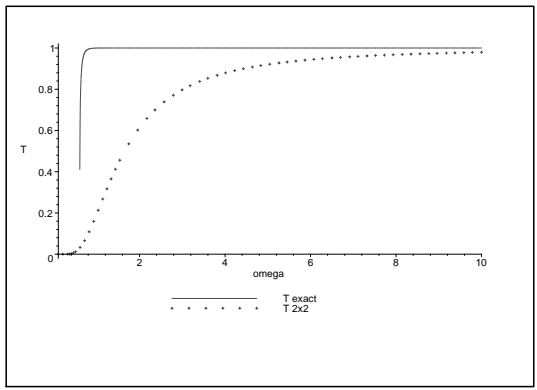,width=2in, angle = -90}}
\vspace*{1pt}
\caption{Dependence of the greybody factor on the energies of the particles emitted from the uncharged dilatonic black holes in (2 + 1) dimensions.}\label{di2oga}
\end{figure}

\noindent Figure \ref{di2oga} shows the greybody factors of the uncharged dilatonic black holes in (2 + 1) dimensions obtained from the $2 \times 2$ transfer matrices [Eq. (\ref{T 2+1})] and from \cite{Fernando} [Eq. (\ref{T 2+1 paper})]. The graph is plotted by setting $m = 0$, $M = 10$, $Q = 0$, and $\Lambda = 0.3$. The graph shows that when the energies of emitted particles increase, the greybody factors also increase. It can be seen that the result derived from the $2 \times 2$ transfer matrices is quite accurate when compared with the exact result. At least, it is, really, a lower bound. Note that the $2 \times 2$ transfer matrices used to obtain the lower bound (\ref{T 2+1}) are much easier than the methods used to obtain the exact result in Eq. (\ref{T 2+1 paper}).

\begin{figure}[pb]
\centerline{\psfig{file=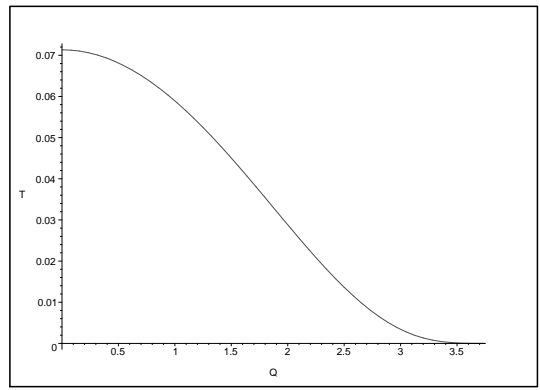,width=2in, angle = -90}}
\vspace*{1pt}
\caption{Dependence of the bound of the greybody factor on the charges for the charged dilatonic black holes in (2 + 1) dimensions.}\label{di21}
\end{figure}

Figure \ref{di21} shows the effect of the charges on the bound of the greybody factor. The graph is plotted by setting $m = 0$, $M = 10$, $\omega = 2$, and $\Lambda = 0.1$. The graph shows that when the charges increases, the bound of the greybody factor decreases. This result is similar to the RN black hole's result; that is the charges behave as good barriers to resist tunneling of uncharged scalar particles.

\begin{figure}[pb]
\centerline{\psfig{file=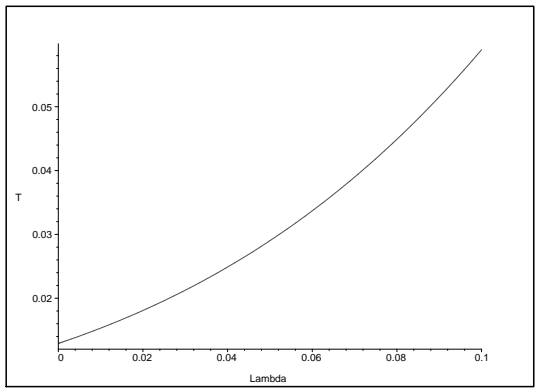,width=2in, angle=-90}}
\vspace*{1pt}
\caption{Dependence of the bound of the greybody factor on the cosmological constant for the charged dilatonic black holes in (2 + 1) dimensions.}\label{di22}
\end{figure}

Figure \ref{di22} shows the effect of the cosmological constant on the bound of the greybody factor. The graph is plotted by setting $m = 0$, $M = 10$, $\omega = 2$, and $Q = 1$. The graph shows that when the value of the cosmological constant increases, the transmission coefficient also increases. That is the cosmological constant makes the gravitational potential produced by the black hole transparent.

\section{The Charged dilatonic black holes in (3 + 1) dimensions}
The Charged dilatonic metric in (3 + 1) dimensions is given by \cite{Kim}
\begin{equation}
ds^{2} = -f(r)dt^{2} + \frac{1}{f(r)}dr^{2} + R^{2}(r)d\Omega^{2},
\end{equation}
where
\begin{equation}
f(r) = 1 - \frac{r_{+}}{r}~~\text{and}~~R^{2}(r) = r^{2}\left(1 - \frac{r_{-}}{r}\right),
\end{equation}
with
\begin{equation}
r_{+} = 2M~~\text{and}~~r_{-} = \frac{Q^{2}}{M}.
\end{equation}
The equation of motion for the radial part is given by
\begin{equation}
\frac{1}{R^{2}(r)}\frac{d}{dr}\left[R^{2}(r)f(r)\frac{du(r)}{dr}\right] + \left[\frac{\omega^{2}}{f(r)} - \frac{l(l + 1)}{R^{2}(r)}\right]u(r) = 0.
\end{equation}
Let
\begin{equation}
dr_{*} = \frac{1}{f(r)}dr,
\end{equation}
then
\begin{eqnarray}
\frac{d^{2}u(r)}{dr_{*}^{2}} + \frac{(r - r_{+})(2r - r_{-})}{r^{2}(r - r_{-})}\frac{du(r)}{dr_{*}}\nn\\
+ \left[\omega^{2} - \frac{l(l + 1)f(r)}{R^{2}(r)}\right]u(r) &=& 0.
\end{eqnarray}

\begin{figure}[pb]
\centerline{\psfig{file=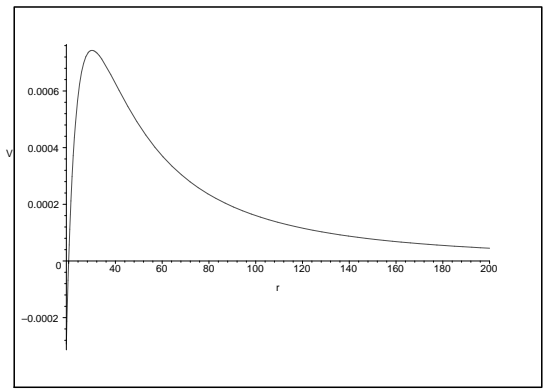,width=2.0in, angle = -90}}
\vspace*{1pt}
\caption{The (3 + 1) charged dilatonic potential with $l = 1$, $Q = 1$, and $M = 10$.}\label{di3 pot}
\end{figure}

\noindent The potential is given by
\begin{equation}
V(r) = \frac{l(l + 1)f(r)}{R^{2}(r)}.
\end{equation}
The (3 + 1) charged dilatonic potential is plotted with $l = 1$, $Q = 1$, and $M = 10$ as shown in Fig. \ref{di3 pot}. The lower bound on the transmission probability for $h = \omega$ is
\begin{eqnarray}
T &\geq& \text{sech}^{2}\left[\frac{1}{2\omega}\int_{-\infty}^{\infty}\frac{l(l + 1)f(r)}{R^{2}(r)}dr_{*}\right]\nn\\
  &=&    \text{sech}^{2}\left[\frac{1}{2\omega}\int_{r_{+}}^{\infty}\frac{l(l + 1)}{R^{2}(r)}dr\right]\nn\\
  &=&    \frac{4\left(2M^{2}\right)^{l(l + 1)M/\omega Q^{2}}\left(2M^{2} - Q^{2}\right)^{l(l + 1)M/\omega Q^{2}}}{\left[\left(2M^{2}\right)^{l(l + 1)M/\omega Q^{2}} + \left(2M^{2} - Q^{2}\right)^{l(l + 1)M/\omega Q^{2}}\right]^{2}}.\nn
\end{eqnarray}

\begin{figure}[pb]
\centerline{\psfig{file=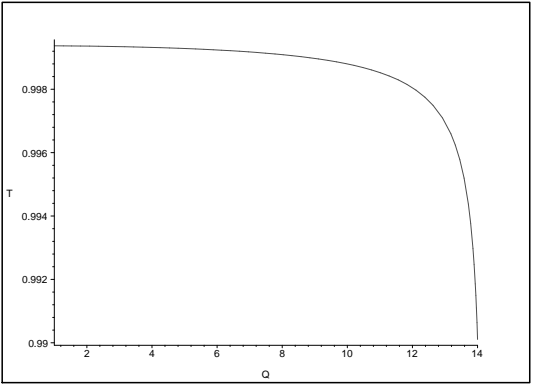,width=2in, angle = -90}}
\vspace*{1pt}
\caption{Dependence of the bound of the greybody factor on the charges for the charged dilatonic black holes in (3 + 1) dimensions.}\label{di31}
\end{figure}

\noindent Figure \ref{di31} shows the effect of the charges on the bound of the greybody factor. The graph is plotted by setting $M = 10$, $\omega = 2$, and $l = 1$. The graph shows that when the charges increases, the bound of the greybody factor decreases. This result is also similar to the Reissner-Nordstr\"{o}m black hole's and the (2 + 1) dimensional charged dilatonic black hole's result. That is the charges behave as good barriers to resist tunneling of uncharged scalar particles.

\section{Conclusions}
The rigorous bounds presented in this paper only work for some potentials. Such potentials have to satisfy $V(\pm\infty) \rightarrow V_{\pm\infty}$. In this paper, the bounds have been applied to various types of black holes.

For the four-dimensional Reissner-Nordstr\"{o}m black holes, the charges act as a good barrier. This can also occur for the charged dilatonic black holes both in (2 + 1) and (3 + 1) dimensions. For the Schwarzschild-Tangherlini black holes, a number of dimensions can shield Hawking radiation.

\begin{acknowledgments}
This research was supported by a grant for the professional development of new academic staff from the Ratchadapisek Somphot Fund at Chulalongkorn University, by Thailand Toray Science Foundation (TTSF), and by the Research Strategic plan program (A1B1), Faculty of Science, Chulalongkorn University. PB was additionally supported by a scholarship from the Royal Government of Thailand. TN was additionally supported by a scholarship from the Development and Promotion of Science and Technology talent project (DPST). TN gives a special thanks to Dr. Auttakit Chatrabuti for his valuable advice. We also thank Prof. Matt Visser for very useful suggestions and comments.
\end{acknowledgments}

\appendix

\section{Schr\"{o}dinger-like equation}

In this appendix we shall present some of the technical details leading to the Schr\"{o}dinger-like equation for scalar modes in the Reissner-Nordstr\"{o}m geometry. The Reissner-Nordstr\"{o}m (RN) metric is given by
\begin{equation}
ds^{2} = -\Delta dt^{2} + \Delta^{-1}dr^{2} + r^{2}d\Omega^{2},\label{RN metric}
\end{equation}
where $d\Omega^{2} = d\theta^{2} + \sin^{2}\theta d\phi^{2}$ and
\begin{equation}
\Delta = 1 - \frac{2GM}{r} + \frac{G\left(Q^{2} + P^{2}\right)}{r^{2}}.\label{delta}
\end{equation}
Therefore, the metric tensor reads
\begin{eqnarray}
\left[g_{\mu\nu}\right] = \left[
                            \begin{array}{cccc}
                              -\Delta & 0           & 0     & 0 \\
                              0       & \Delta^{-1} & 0     & 0 \\
                              0       & 0           & r^{2} & 0 \\
                              0       & 0           & 0     & r^{2}\sin^{2}\theta \\
                            \end{array}
                          \right]
\end{eqnarray}
and the inverse metric is given by
\begin{eqnarray}
\left[g^{\mu\nu}\right] = \left[
                            \begin{array}{cccc}
                              -\Delta^{-1} & 0      & 0      & 0 \\
                              0            & \Delta & 0      & 0 \\
                              0            & 0      & r^{-2} & 0 \\
                              0            & 0      & 0      & r^{-2}\sin^{-2}\theta \\
                            \end{array}
                          \right].
\end{eqnarray}
Consider a massless uncharged scalar field in this background. Its equation of motion is the Klein-Gordon equation
\begin{equation}
\nabla^{\mu}\partial_{\mu}\Phi = 0.\label{KG}
\end{equation}
Since the covariant derivative of a one-form is given by
\begin{equation}
\nabla_{\nu}\omega_{\mu} = \partial_{\nu}\omega_{\mu} - \Gamma^{\lambda}_{\nu\mu}\omega_{\lambda},
\end{equation}
where the Christoffel symbol is defined by
\begin{equation}
\Gamma^{\lambda}_{\nu\mu} = \frac{1}{2}g^{\lambda\rho}\left(\partial_{\nu}g_{\mu\rho} + \partial_{\mu}g_{\rho\nu} - \partial_{\rho}g_{\nu\mu}\right),
\end{equation}
we obtain
\begin{equation}
\nabla^{\mu}\partial_{\mu}\Phi = g^{\mu\nu}\partial_{\mu}\partial_{\nu}\Phi - g^{\mu\nu}\Gamma^{\lambda}_{\mu\nu}\partial_{\lambda}\Phi.\label{left KG}
\end{equation}
Now, we compute the nonzero Christoffel symbols. We derive
\begin{eqnarray}
\Gamma^{1}_{00} &=& \frac{1}{2}\Delta d_{r}\Delta,\nn\\
\Gamma^{1}_{11} &=& -\frac{1}{2}\Delta^{-1}d_{r}\Delta,\nn\\
\Gamma^{1}_{22} &=& -\Delta r,\nn\\
\Gamma^{1}_{33} &=& -\Delta r\sin^{2}\theta,\nn\\
\Gamma^{2}_{33} &=& -\sin\theta\cos\theta.\nn
\end{eqnarray}
Putting them in Eq. (\ref{left KG}), we obtain
\begin{eqnarray}
\nabla^{\mu}\partial_{\mu}\Phi &=& -\Delta^{-1}\partial_{t}^{2}\Phi + \Delta\partial_{r}^{2}\Phi + \frac{1}{r^{2}}\partial_{\theta}^{2}\Phi + \frac{1}{r^{2}\sin^{2}\theta}\partial_{\phi}^{2}\Phi\nn\\
                               && + d_{r}\Delta\partial_{r}\Phi + \frac{2}{r}\Delta\partial_{r}\Phi + \frac{\cot\theta}{r^{2}}\partial_{\theta}\Phi.
\end{eqnarray}
Therefore, the Klein-Gordon equation (\ref{KG}) becomes
\begin{eqnarray}
-\Delta^{-1}\partial_{t}^{2}\Phi + \Delta\partial_{r}^{2}\Phi + \frac{1}{r^{2}}\partial_{\theta}^{2}\Phi\nn\\
+ \frac{1}{r^{2}\sin^{2}\theta}\partial_{\phi}^{2}\Phi + d_{r}\Delta\partial_{r}\Phi\nn\\
+ \frac{2}{r}\Delta\partial_{r}\Phi + \frac{\cot\theta}{r^{2}}\partial_{\theta}\Phi &=& 0.\label{full KG}
\end{eqnarray}
Let
\begin{equation}
\Phi = U(t, r)Y(\theta, \phi).
\end{equation}
Putting it into Eq. (\ref{full KG}), we derive
\begin{eqnarray}
\frac{r^{2}\Delta^{-1}}{U}\partial_{t}^{2}U - \frac{r^{2}\Delta}{U}\partial_{r}^{2}U\nn\\
- \frac{r^{2}d_{r}\Delta}{U}\partial_{r}U- \frac{2r\Delta}{U}\partial_{r}U &=& \frac{1}{Y}\partial_{\theta}^{2}Y + \frac{1}{Y\sin^{2}\theta}\partial_{\phi}^{2}Y\nn\\
                                  && + \frac{\cot\theta}{Y}\partial_{\theta}Y.
\end{eqnarray}
We choose each side of the above equation to be equal to a constant $-k^{2}$. The equation for $U(t, r)$ is
\begin{equation}
-\Delta^{-1}\partial_{t}^{2}U + \Delta\partial_{r}^{2}U + d_{r}\Delta\partial_{r}U + \frac{2}{r}\Delta\partial_{r}U = \frac{k^{2}}{r^{2}}U,\label{U}
\end{equation}
while the equation for $Y(\theta, \phi)$ is
\begin{equation}
\partial_{\theta}^{2}Y + \frac{1}{\sin^{2}\theta}\partial_{\phi}^{2}Y + \cot\theta\partial_{\theta}Y = -k^{2}Y.
\end{equation}
Proceeding further, we obtain
\begin{equation}
\frac{1}{\sin\theta}\partial_{\theta}\left[\sin\theta(\partial_{\theta}Y)\right] + \frac{1}{\sin^{2}\theta}\partial_{\phi}^{2}Y + k^{2}Y = 0.
\end{equation}
We find that $k^{2} = l(l + 1)$. Then, Eq. (\ref{U}) becomes
\begin{equation}
-\Delta^{-1}\partial_{t}^{2}U + \Delta\partial_{r}^{2}U + d_{r}\Delta\partial_{r}U + \frac{2}{r}\Delta\partial_{r}U = \frac{l(l + 1)}{r^{2}}U.
\end{equation}
Let
\begin{equation}
U = T(t)\varphi(r),
\end{equation}
then
\begin{equation}
\frac{1}{T}d_{t}^{2}T - \frac{\Delta^{2}}{\varphi}d_{r}^{2}\varphi - \frac{\Delta d_{r}\Delta}{\varphi}d_{r}\varphi - \frac{2\Delta^{2}}{r\varphi}d_{r}\varphi = -\frac{l(l + 1)\Delta}{r^{2}}.\label{Tphi}
\end{equation}
Let
\begin{equation}
\frac{1}{T}d_{t}^{2}T = -\omega^{2},
\end{equation}
then Eq. (\ref{Tphi}) becomes
\begin{equation}
\Delta^{2}d_{r}^{2}\varphi + \Delta d_{r}\Delta d_{r}\varphi + \frac{2\Delta^{2}}{r}d_{r}\varphi = \left[\frac{l(l + 1)\Delta}{r^{2}} - \omega^{2}\right]\varphi.\label{varphi}
\end{equation}
Now, we make change of variable
\begin{equation}
\varphi(r) = \frac{1}{r}\psi(r).
\end{equation}
Putting it into Eq. (\ref{varphi}), we derive
\begin{equation}
\Delta^{2}d_{r}^{2}\psi + \Delta d_{r}\Delta d_{r}\psi + \omega^{2}\psi = \left[\frac{l(l + 1)\Delta}{r^{2}} + \frac{\Delta d_{r}\Delta}{r}\right]\psi.\label{psi}
\end{equation}
We define
\begin{equation}
d_{r_{*}}^{2} \equiv \Delta^{2}d_{r}^{2} + \Delta d_{r}\Delta d_{r}.\label{x def}
\end{equation}
Therefore,
\begin{equation}
d_{r}r_{*} = \Delta^{-1},\label{dx}
\end{equation}
then we derive
\begin{equation}
r_{*} = \int\Delta^{-1}dr.
\end{equation}
From Eq. (\ref{delta}), we can write $r_{*}$ in terms of $r$
\begin{eqnarray}
r_{*} = \left\{
      \begin{array}{ll}
        r + GM\ln\left|u^{2} - A^{2}\right|                                   & \\
        + \dfrac{G^{2}M^{2} + A^{2}}{2A}\ln\left|\dfrac{u - A}{u + A}\right|, & GM^{2} > Q^{2} + P^{2} \\
        r + GM\ln\left|u^{2} + B^{2}\right|                                   & \\
        + \dfrac{G^{2}M^{2} - B^{2}}{B}\arctan\dfrac{u}{B},                   & GM^{2} < Q^{2} + P^{2} \\
        r + GM\ln\left|u^{2}\right| - \dfrac{G^{2}M^{2}}{u},                  & GM^{2} = Q^{2} + P^{2} \\
      \end{array}
    \right.,\nn
\end{eqnarray}
where
\begin{eqnarray}
u     &=& r - GM\nn\\
A^{2} &=& G^{2}M^{2} - G\left(Q^{2} + P^{2}\right)\\
B^{2} &=& -G^{2}M^{2} + G\left(Q^{2} + P^{2}\right).\nn
\end{eqnarray}
With this coordinate $r_{*}$ defined in Eq. (\ref{x def}), Eq. (\ref{psi}) becomes
\begin{equation}
\frac{d^{2}\psi}{dr_{*}^{2}} + \left[\omega^{2} - V(r)\right]\psi = 0,
\end{equation}
where
\begin{equation}
V(r) = \frac{l(l + 1)\Delta}{r^{2}} + \frac{\Delta\partial_{r}\Delta}{r}.
\end{equation}

\section{Greybody factor from the WKB approximation; Derivation of Eq. (\ref{Twkb RN})}
By using the WKB method, the approximate transmission coefficient is given by \cite{Lange}
\begin{equation}
T \approx T_{\text{WKB}} = \exp\left[-\frac{2}{\hbar}\text{Im}\int_{a}^{b}p(x)dx\right],\label{T}
\end{equation}
where
\begin{equation}
p(x) = \sqrt{2m[E - V(x)]}.
\end{equation}
In particular, we want to compute
\begin{equation}
\int_{r_{\text{in}}}^{r_{\text{out}}}p_{r}dr.\nn
\end{equation}
The radial momentum can be written as an integral
\begin{equation}
\int_{r_{\text{in}}}^{r_{\text{out}}}p_{r}dr = \int_{r_{\text{in}}}^{r_{\text{out}}}\int_{0}^{p_{r}}dp'_{r}dr.
\end{equation}
From the Hamilton equation
\begin{equation}
\frac{dH}{dp_{r}} = \dot{r},
\end{equation}
the above integral becomes
\begin{equation}
\int_{r_{\text{in}}}^{r_{\text{out}}}p_{r}dr = \int_{r_{\text{in}}}^{r_{\text{out}}}\int_{M}^{M - \omega}\frac{dH}{\dot{r}}dr.
\end{equation}
We change the variable $H$ to $\omega'$
\begin{equation}
\int_{r_{\text{in}}}^{r_{\text{out}}}p_{r}dr = -\int_{r_{\text{in}}}^{r_{\text{out}}}\int_{0}^{\omega}\frac{d\omega'}{\dot{r}}dr.\label{int d omega}
\end{equation}
We have to know $\dot{r}$. Starting from the Reissner-Nordstr\"{o}m metric in Eq. (\ref{RN metric})
\begin{equation}
ds^{2} = -\Delta dt_{RN}^{2} + \Delta^{-1}dr^{2} + r^{2}d\Omega^{2},
\end{equation}
we shift the Reissner-Nordstr\"{o}m time $t_{RN}$ by a function of $r$ to avoid the singularities
\begin{eqnarray}
t_{RN}      &=& t + f(r)\nn\\
dt_{RN}     &=& dt + f'(r)dr\nn\\
dt_{RN}^{2} &=& dt^{2} + 2f'(r)dtdr + [f'(r)]^{2}dr^{2}.\nn
\end{eqnarray}
Therefore,
\begin{equation}
ds^{2} = -\Delta dt^{2} - 2\Delta f'(r)dtdr - \Delta[f'(r)]^{2}dr^{2} + \Delta^{-1}dr^{2} + r^{2}d\Omega^{2}.\label{new metric}
\end{equation}
We choose $f(r)$ such that the coefficient of $dr^{2}$ is equal to one
\begin{eqnarray}
\Delta f'(r) &=& \pm\sqrt{\frac{2GM}{r} - \frac{G\left(Q^{2} + P^{2}\right)}{r^{2}}}\nn\\
             &=& \pm\sqrt{1 - \Delta}.
\end{eqnarray}
Putting it in Eq. (\ref{new metric}), the new metric can be written as
\begin{equation}
ds^{2} = -\Delta dt^{2} + 2\sqrt{1 - \Delta}dtdr + dr^{2} + r^{2}d\Omega^{2}.
\end{equation}
The radial null geodesics can be found by
\begin{equation}
0 = ds^{2} = -\Delta dt^{2} + 2\sqrt{1 - \Delta}dtdr + dr^{2},
\end{equation}
leading to
\begin{eqnarray}
\dot{r} &=& \left\{
                                                       \begin{array}{cc}
                                                         1 - \sqrt{1 - \Delta}  &  \\
                                                         -1 - \sqrt{1 - \Delta} &  \\
                                                       \end{array}
                                                     \right..
\end{eqnarray}
Therefore, integral (\ref{int d omega}) becomes
\begin{equation}
\int_{r_{\text{in}}}^{r_{\text{out}}}p_{r}dr = -\int_{r_{\text{in}}}^{r_{\text{out}}}\int_{0}^{\omega}\frac{r}{r - x}d\omega'dr,
\end{equation}
where
\begin{equation}
x = \sqrt{2G\left(M - \omega'\right)r - G\left(Q^{2} + P^{2}\right)}.
\end{equation}
Thus, $dx = -(Gr/x)d\omega'$ and we obtain
\begin{eqnarray}
\int_{r_{\text{in}}}^{r_{\text{out}}}p_{r}dr &=& \int_{r_{\text{in}}}^{r_{\text{out}}}\int_{\sqrt{2GMr - G\left(Q^{2} + P^{2}\right)}}^{\sqrt{2G\left(M - \omega\right)r - G\left(Q^{2} + P^{2}\right)}}\frac{r}{r - x}\frac{x}{Gr}dxdr\nn\\
                                             &=& \pi i\left[2G\omega\left(M - \frac{\omega}{2}\right)\right.\nn\\
                                             && - (M - \omega)\sqrt{G^{2}\left(M - \omega\right)^{2} - G\left(Q^{2} + P^{2}\right)}\nn\\
                                             && \left. + M\sqrt{G^{2}M^{2} - G\left(Q^{2} + P^{2}\right)}\right].
\end{eqnarray}
Therefore, from Eq. (\ref{T})
\begin{eqnarray}
T &\approx& T_{\text{WKB}}\nn\\
  &=&       \exp\left[-\frac{2\pi}{\hbar}\left\{2G\omega\left(M - \frac{\omega}{2}\right)\right.\right.\nn\\
  && - (M - \omega)\sqrt{G^{2}\left(M - \omega\right)^{2} - G\left(Q^{2} + P^{2}\right)}\nn\\
  && \left.\left. + M\sqrt{G^{2}M^{2} - G\left(Q^{2} + P^{2}\right)}\right\}\right].
\end{eqnarray}

\end{document}